\documentclass{elsart}
\usepackage{amsfonts}
\usepackage{graphicx}
\usepackage{dcolumn}

\begin{document}
\newcommand{\be}{\begin{equation}}
\newcommand{\ee}{\end{equation}}
\newcommand{\ber}{\begin{eqnarray}}
\newcommand{\eer}{\end{eqnarray}}
\newcommand{\mean}[1]{\left\langle #1 \right\rangle}
\newcommand{\abs}[1]{\left| #1 \right|}
\newcommand{\set}[1]{\left\{ #1 \right\}}
\newcommand{\la}{\langle}
\newcommand{\ra}{\rangle}
\newcommand{\lb}{\left(}
\newcommand{\rb}{\right)}
\newcommand{\norm}[1]{\left\|#1\right\|}
\newcommand{\RA}{\rightarrow}
\newcommand{\tet}{\vartheta}
\newcommand{\eps}{\varepsilon}
\newcommand{\tNN}{\tilde{\mathbf{X}}_n^{NN}}
\newcommand{\NN}{\mathbf{X}_n^{NN}}

\begin{frontmatter}

\title{Investment strategy due to the minimization of  portfolio noise level  by
 observations of coarse-grained entropy}
\author{Krzysztof Urbanowicz and Janusz A. Ho{\l}yst}
\address{Faculty of Physics and Center of Excellence for Complex
Systems Research \\Warsaw University of Technology \\
Koszykowa 75, PL--00-662 Warsaw, Poland}

\date{\today}

\begin{abstract}
Using a recently developed method of noise level estimation that
makes use of  properties of the coarse grained-entropy we have
analyzed the noise level for the Dow Jones index and a few stocks
from the New York Stock Exchange. We have found that the noise
level ranges from $40$ to $80$ percent of the signal variance. The
condition of a minimal noise level has been applied to construct
optimal portfolios from selected shares. We show that
implementation of a corresponding threshold investment strategy
leads to positive returns for historical data.
\end{abstract}

\end{frontmatter}

\section{Introduction}
\par Although it is a common believe that the stock
market behaviour is driven by  stochastic processes
\cite{voit,Buchound,Mantegna} it is difficult to separate
stochastic and deterministic  components   of market dynamics. In
fact the deterministic fraction  follows  usually from  nonlinear
effects and can possess  a non-periodic or even chaotic
characteristic \cite{Peters,Holyst}. The aim of this paper is to
study the level of determinism in time series coming from stock
market. We will show that our noise level analysis can be useful
for portfolio optimization.
\par We employ here a  method of noise-level estimation that has been
 described in details in \cite{urbanowicz}.
 The method is quite universal and it  is valid even for high noise
 levels.
It makes use of the functional dependence of coarse-grained
correlation entropy $K_2(\eps)$ \cite{kantzschreiber} on the
threshold parameter $\eps$. Since the function $K_2(\eps)$ depends
in a characteristic way on the noise standard deviation $\sigma$
thus  one can estimate the noise level $\sigma$ observing the
dependence  $K_2(\eps)$. The validity of our method has been
verified by applying it for the noise level estimation in several
chaotic models \cite{kantzschreiber} and for the Chua electronic
circuit contaminated by noise. The method distinguishes a noise
appearing due to the presence of a stochastic process from a
non-periodic {\it deterministic} behaviour (including the
deterministic chaos). Analytic calculations justifying our method
have been developed for the gaussian noise added to the observed
deterministic variable. It has been also checked in numerical
experiments  that the method works properly for a uniform noise
distribution and at least for some models with dynamical noise
corresponding to the Langevine equation \cite{urbanowicz}.
\section{Calculations of noise level in  stock market data}

We define the noise level as the ratio of standard deviation of
noise $\sigma$ to the standard deviation of data $\sigma_{data}$
\begin{equation} NTS=\frac{\sigma}{\sigma_{data}}\end{equation}

\begin{figure}
\begin{center}
\includegraphics[scale=0.5,angle=-90]{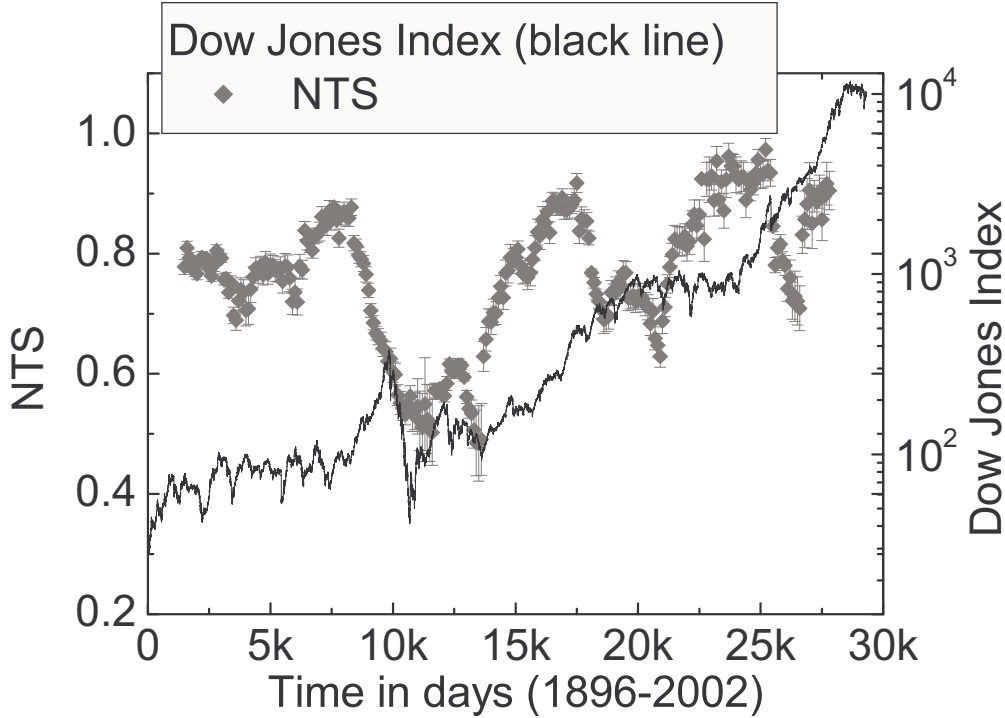}
\end{center}
\caption{\label{DJIANTS} The noise level $NTS$ calculated for Dow
Jones index (1896-2002).}
\end{figure}

Using this definition and our noise level estimation method
\cite{urbanowicz} we have analyzed the noise level in data
recorded at the New York Stock Exchange (NYSE). Let us consider
logarithmic daily returns for the  Dow Jones Industrial Average
(DJIA)

    \be x_i=\ln \lb \frac{P_{i}}{P_{i-1}}\rb.\ee
Fig.~\ref{DJIANTS} presents the plot of the noise level $NTS$ for
a corresponding time series  $x_i$ where values of $NTS$  have
been calculated as a function of a trading day. The noise level
has been  determined in windows of the length $3000$ days and is
pointed in the middle of every  window. As one can see the level
of noise ranges from $60\%$ to $90\%$ what makes any point to
point forecasting impossible. We should mention that since the
relative noise variance is  $NTS^2$, thus in our case the noise
variance is $40-80\%$ of  the data variance. It follows that there
are time periods when  the percent of an unknown deterministic
part approaches the level $60\%$ of the signal.
\par
Similar estimations of the noise level  have been performed for
selected stocks of the NYSE. Results for the mean values of
corresponding NTS parameters are presented in the
Table~\ref{tab:tabNTSspol}. As one can expect  the noise level of
a single stock is much larger than for the DJIA. This is because
deterministic parts of different stock prices are usually positive
correlated what is less common  for stochastic components.

\par The crucial point for our  investment strategy are
correlations between a temporary value of  noise level and a
temporary value of price changes. We have found that for the
majority of considered stocks  the correlation coefficient $\rho$
is much larger than zero (see the Table~\ref{tab:tabNTSspol}) in
the time period when  trends of these stock were negative. A
negative correlation coefficient has been observed for one share
with a positive share. Following these observations we have formed
the following heuristic rule:  temporary price changes are mostly
consistent with the trend for a small noise level but they are
frequently opposed to the trend for a large noise level (the noise
level should be measured locally). One can say that when price
changes are more stochastic investors are more disoriented than
for a more deterministic price motion and they more frequently
trade against the general trend.
\begin{table}
\begin{center}
\caption{\label{tab:tabNTSspol} Stocks recorded at the NYSE for
the period 01.01.1999-31.12.2000. Table of the values of the noise
levels $NTS$, correlation coefficient  $\rho$ between prices
changes and the noise levels and returns in the studied period.}

\begin{tabular}{lccc}
\hline
 Stock& $NTS$& $\rho$& returns at the period\\
\hline
 Apple (Ap) & $77\%$& $0.53$& $-63\%$\\
 Bank of America (Boa)& $93\%$& $0.22$& $-24\%$\\
 Boeing (Bg)& $94\%$& $-0.45$& $98\%$\\
 Cisco (Ci)& $85\%$& $0.57$& $-59\%$\\
 Compaq (Cq)& $93\%$& $-0.09$& $-66\%$\\
 Ford (Fo)& $91\%$& $-0.16$&$-60\%$\\
 General Electric (Ge)& $86\%$& $0.75$&$-53\%$\\
 General Motors (Gm)& $91\%$& $0.44$&$-28\%$\\
 Ibm (Ibm)& $75\%$ & $0.023$ & $-55\%$\\
 Mcdonald (Md)& $93\%$& $0.2$&$-57\%$\\
 Texas Instrument (Te)& $76\%$& $0.77$&$-45\%$\\
\hline
\end{tabular}
\end{center}
\end{table}

\section{Investment strategy}
\par Using the fact that the level of noise is correlated with the
stock price changes, one can create a portfolio which can maximize
the profit. In the first step we construct a portfolio with the
minimal value of the stochastic variable. We assume that one can
do this by maximization of the following quantity: \be
\mathcal{B}=\sum\limits_{i=1}^{M}\sum\limits_{j=1}^M p_i p_j
\frac{\sigma_{i,D}}{\sigma_{i}}\frac{\sigma_{j,D}}{\sigma_{j}}\rho_{i,j}=max.
\ee where $\sigma_{i,D}$ is a standard deviation of deterministic
part of the stock $i$, $\sigma_i$ is the standard deviation of the
noise in this stock and $\rho_{i,j}$ is the correlation between
deterministic parts  of stocks $i$ and $j$. The maximal value of
$\mathcal{B}$ can be performed  with the help of the steepest
descent method by changing the variables $p_i$ and keeping the
normalization constraint  $\sum_{i=1}^M p_i = 1$.
\par Now let us define our  investment
strategy  as follows: if the past trend of portfolio is positive
$m_p>0$ and the noise level is small ($NTS_p<NTS_{treshhold}$) we
invest in the calculated portfolio. We invest also in the
portfolio when it is more stochastic ($NTS_p>NTS_{treshhold}$) but
its trend is negative $m_p<0$. We invest against  the portfolio in
the remaining two cases.
\par Table~\ref{tab:port811} presents the values $P_{p}$ for a few
portfolios at the end of a trading period  when the above
investment strategy has been used. The results are compared to
mean values of the prices of stocks $P_{m}$ at the same moment
  and a relative profit of  our investment strategy is shown: $(P_p/P_m-1)\cdot 100\%$.
 To get the proper normalization we set  the values  $P_{p}$ and $P_{m}$  to one
  at the beginning of the trading period. Although  the above analysis
  gives very promising results one should mention that   all
  commissions costs have been omitted and what is more crucial we have assumed an
  unlimited  possibility of short-sellings. As result our portfolios are very risky.
  When one limits the possible short-selling level the risk and returns are lower.

  In the Table~\ref{tab:portsum} the results are summarized  for
 several studied portfolios. We have found  that $62\%$ of portfolios
 had positive returns even that for the considered time period almost all single stock
  returns were negative (see Table~\ref{tab:tabNTSspol}).
\begin{table}
\begin{center}
\caption{\label{tab:port811} The value of optimized portfolio
$P_p$ at the end of studied period (01.01.1999-31.12.2000) in
comparison to behavior of mean prices $P_m$.}

\begin{tabular}{lccc}
\hline
 Stocks& $P_m$ & $P_p$ & $(P_p/P_m-1)\cdot 100\%$\\
\hline
 Ap, Bg, Cq, Ge, Ibm, Md, Boa, Ci & $0.58$& $3.41$& $487\%$\\
 Ap, Boa, Bg, Ci, Cq, Fo, Ge, Gm & $0.56$& $6.65$& $1087\%$\\
 Bg, Ci, Cq, Fo, Ge, Gm, Ibm, Te& $0.60$& $1.01$& $68\%$\\
 Boa, Bg, Ci, Cq, Fo, Ge, Gm, Ibm& $0.64$& $0.48$&$-25\%$\\
 Ci, Cq, Fo, Ge, Gm, Ibm, Te, Md& $0.55$& $1.86$& $238\%$\\
 Md, Ibm, Cq, Te, Boa, Fo, Ap, Ge& $0.48$&$8.08$& $1583\%$\\
 all studied stocks & $0.56$ & $8.45$ & $1408\%$\\
 \hline
\end{tabular}
\end{center}
\end{table}

\begin{table}
\begin{center}
\caption{\label{tab:portsum}  Summary of the portfolio
optimization. Analysis was done for the period
01.01.1999-31.12.2000.}

\begin{tabular}{lc}
\hline
 No of elaborated portfolios & $55$ \\
 Portfolios with negative returns& $38\%$ \\
 Portfolios with positive returns& $62\%$ \\
 Portfolios with returns smaller than mean& $29\%$ \\
 Portfolios with returns larger than mean& $71\%$ \\
\hline
\end{tabular}
\end{center}
\end{table}
\section{Conclusions}
\par In conclusion we have found that the deterministic part  of stock market
data at NYSE is in the range  $20-60\%$ of the data variance. The
estimation of noise level can be useful for portfolio
optimization. The resulting investment strategy gives in average
positive returns.

\section{Acknowledgement}
This work has been partially supported by the KBN Grant 2 P03B 032
24 (KU) and by a special Grant {\it Dynamics of Complex Systems}
of the Warsaw University of Technology (JAH).

\end{document}